\documentclass[10pt]{article}
\hoffset= - 2.5cm
\usepackage{amssymb,graphicx}
\usepackage{epstopdf}
\usepackage{amsmath,amsfonts}
\usepackage{epsfig}
\usepackage[dvipsnames]{xcolor}
\usepackage{hyperref}
\definecolor{linkcolor}{HTML}{799B03}
\definecolor{urlcolor}{HTML}{799B03}
\hypersetup{pdfstartview=FitH,linkcolor=linkcolor,urlcolor=urlcolor,colorlinks=true}
\newcommand*{\cG}{\mathcal{G}}
\newcommand*{\cF}{\mathcal{F}}

\newcommand*{\cD}{\mathcal{D}}
\newcommand*{\cT}{\mathcal{T}}
\newcommand*{\cL}{\mathcal{L}}
\newcommand*{\cS}{\mathcal{S}}
\def\tz{\tilde{\zeta}}
\def\[{\begin{equation}}
\def\]{\end{equation}}

\begin{document}

\begin{center}
  {\LARGE \bf  Non-singular cosmological scenarios in scalar-tensor theories and their stability: a review}

\vspace{10pt}

{\bf  An essay in honor of V.A. Rubakov on the occasion of his 70th birthday}

\vspace{20pt}
S. Mironov$^{a,b,d}$\footnote{sa.mironov\_1@physics.msu.ru},
V. Volkova$^{a,c}$\footnote{volkova.viktoriya@physics.msu.ru}
\renewcommand*{\thefootnote}{\arabic{footnote}}
\vspace{15pt}

$^a$\textit{Institute for Nuclear Research of the Russian Academy of Sciences, 117312 Moscow, Russia}

$^b$\textit{Institute for Theoretical and Mathematical Physics, MSU, 119991 Moscow, Russia}

$^c$\textit{Department of Particle Physics and Cosmology, Physics Faculty, MSU, 119991 Moscow, Russia}

$^d$\textit{NRC "Kurchatov Institute", 123182, Moscow, Russia}
\end{center}

\vspace{5pt}

\begin{abstract}
This article gives a concise overview of the development and current status of studies on healthy models of the early Universe without an initial singularity, namely the cosmological bounce and Genesis scenarios, constructed within a broad class of scalar-tensor theories, specifically Horndeski theories and their generalizations. The review focuses on the topics related to linear stability at the perturbation level over the non-singular background solutions: 1) the no-go theorem, valid for non-singular cosmologies within Horndeski theory, 2) the updates on possible approaches to evade the no-go theorem, 3) the role of disformal transformations relating the Horndeski subclasses with the generalized theories like DHOST, 4) the effects on stability caused by additional matter coupling and potential emergence of superluminal perturbation modes in the multi-component setting.

\end{abstract}

%%%%%%%%%%%%%%%%%%%%%%%%%%%%%%%%%%%%%%%%%%%%%%%%%%%%%%%%%%%%%%%%%%%%%%
\section{Introduction}

There is much that is known and much that is unknown about the physics of the early Universe, and the further back in time we go, the less we can say about the Universe with any degree of confidence. 
While our knowledge of the hot big bang stages, starting with the fractions of a second of the Universe's lifetime, is quite solid from both a theoretical and an observational point of view, we delve into yet purely theoretical domain when the temperature climbs up to hundreds MeV. 

The common standpoint today is that the conventional hot big bang cosmological stages were preceded by the epoch(s) that have to resolve the standard set of the big bang problems, namely, the horizon puzzle, the spatial flatness problem, the entropy dilemma and the issue of primordial density fluctuations. The concept of inflation~\cite{Starobinsky:1980te,Guth:1980zm} elegantly resolves these problems, but the inflationary scenario itself faces, among other issues~\cite{Brandenberger:1999sw,Turok:2002yq}, a serious challenge of geodesic incompleteness in the past, i.e. it cannot solve the initial singularity problem~\cite{Borde:1996pt,Borde:2001nh}. Hence, the inflationary epoch cannot yield the complete discription of the very early Universe
\footnote{However, the inflationary concept may be considered from a somewhat broader perspective. In particular, Refs.~\cite{Lesnefsky:2022fen,Easson:2024uxe,Easson:2024fzn} consider inflationary models which are in a way similar to the Genesis scenario and feature an asymptotically vanishing Hubble parameter at early times. This kind of cosmological dynamics resolves the standard problem of geodesic incompleteness in the past. Importantly, this type of inflationary models requires violation of the Null Energy Conditions (see the discussion in Sec.~\ref{sec:NEC}) just like 
the bouncing and Genesis models do.}. 

One of the alternative or possibly complementary cosmological models, that addresses the initial singularity problem, is the non-singular bouncing scenario, where the Universe initially undergoes a contraction phase, and then transits into the currently expanding phase through a moment of a ''bounce'' characterized by a finite value of the scale factor (see~\cite{Lehners:2008vx,Novello:2008ra,Cai:2014bea,Battefeld:2014uga} for reviews). Another non-singular model is the Universe starting off with the cosmological Genesis, where the Universe starts its expansion from the asymptotically empty Minkowski space~\cite{Creminelli:2006xe,Creminelli:2010ba}. Both these scenarios share the feature of relying on the non-trivial dynamics, which requires the Hubble parameter to grow. Indeed, the bouncing concept implies, that the Hubble rate transits from negative to positive values at the moment of the bounce; the Genesis scenario implies that on the onset the Hubble parameter is vanishing and starts growing when the expansion launches. Such behaviour of the Hubble rate is impossible within the framework of General Relativity with a conventional matter, respecting all the energy conditions. Hence, the non-standard dynamics of both scenarios comes at a price of introducing an exotic matter component, which would violate the Null Energy Condition (NEC) and allow for avoiding the singularity without abandoning General Relativity or making use of space curvature. The violation of the NEC often leads to the appearance of pathological degrees of freedom among perturbations over a homogeneous background (see~\cite{RubakovNEC} for review), which makes it challenging to construct a viable model with a bouncing or Genesis stage, where the instabilities are absent.

In this mini review we focus on Horndeski theories and their generalizations as a suitable setting for the healthy NEC-violation, which makes this wide class of scalar-tensor theories to be a promising framework for constructing viable non-singular cosmologies. The main emphasis is put on analysing and verifying stability of a non-singular solutions at the linearized level: we discuss in details the necessary criteria which enable one to ensure that all types of instabilities are absent over the background throughout the entire evolution of the model. We discuss the potential troubles for the stability related to coupling additional matter components to the scalar-tensor setup. We also address some recent developments in the field and e.g. briefly discuss the new classes of theories which further generalize Horndeski family. 

This review has the following structure. In Sec.~\ref{sec:NEC} we discuss in details the NEC and its implications for the non-singular cosmologies and in Sec.~\ref{sec:STT} we briefly review the structure of Horndeski theory and its generalizations. The explicit analysis of perturbations' behaviour over the cosmological background in beyond Horndeski (or, equivalently, GLPV) theory is given in Sec.~\ref{sec:stability} and provides the base for the stability no-go theorem proved for Horndeski theories in Sec.~\ref{sec:theorem}. In Sec.~\ref{sec:strong_coupling}-\ref{sec:BH} we discuss different ways to evade the no-go theorem and analyse the role of disformal transformsions in this context in Sec.~\ref{sec:no-go_disformal}. We address the issues whether the matter coupling cause troubles for the stability in the generalized Horndeski theory in Sec.~\ref{sec:matter_coupling} and also separately analyse the potential problem with emergent superluminal modes upon coupling of the extra scalar field in Sec.~\ref{sec:matter_coupling_superlum}.
We briefly conclude in Sec.~\ref{sec:conclusion}.

%%%%%%%%%%%%%%%%%%%%%%%%%%%%%%%%%%%
%%%%%%%%%%%%%%%%%%%%%%%%%%%%%%%%%%%
\subsection{Null Energy Condition and its violation}
\label{sec:NEC}

Energy conditions play a significant role in General Relativity (GR) as they
provide natural restrictions for the energy-momentum tensor $T_{\mu\nu}$ of a physical system. Among other energy conditions, 
the Null Energy Condition (NEC), which states that
\[
\label{eq:NEC}
T_{\mu\nu}n^{\mu}n^{\nu} \geq 0,
\] 
for any null vector $n^{\mu}$ satisfying $g_{\mu\nu}n^{\mu}n^{\nu}=0$,
plays the special role.
Despite being the weakest of all, the NEC is particularly robust and fundamental, as it cannot be violated by adjusting the vacuum energy contribution and, hence, unambiguously refers to the properties of matter.
For an isotropic fluid with pressure $p$ and positive energy density $\rho$, the NEC amounts to
\[
\label{eq:NEC_perfectfluid}
\omega = \frac{p}{\rho} \geq -1.
\]
Moreover, violating the NEC has been closely associated with catastrophic consequences from the stability point of view, since NEC violation gave rise
to various pathologies including ghosts, gradient instabilities 
(exponentially growing modes with arbitrarily short wavelengths), superluminality, etc.~\cite{Dubovsky:2005xd}.

Another aspect, which makes the NEC stand out, is the validity of the Penrose singularity theorem~\cite{Penrose:1964wq}, where the NEC is assumed to be satisfied. In particular, the Penrose theorem almost precludes the non-singular bouncing scenario within GR. The latter can be seen 
immediately from
the linear combination of $(00)$- and $(ij)$-components of Einstein equations written for a homogeneous, isotropic Universe with a Friedmann–Lemaıtre–Robertson–Walker (FLRW) metric~\cite{RubakovNEC}:
\[
\label{eq:dotH}
\dot{H} = - 4\pi G (\rho + p) + \frac{\kappa}{a^2},
\]
where $a$ is a scale factor, $H \equiv \dot{a}/{a}$ is a Hubble parameter and $\kappa$ is a curvature parameter ($\kappa = +1$ for the closed Universe, $\kappa = -1$ for the open Universe and 
$\kappa = 0$ for a flat Euclidean 3D space). 
Indeed, $\dot{H} \leq 0$ in a spatially-flat Universe ($\kappa=0$) or the open Universe ($\kappa = -1$)
since the NEC implies that $\rho+p \geq 0$, see eq.~\eqref{eq:NEC_perfectfluid}. Hence, 
the contraction ($H < 0$) cannot evolve into expansion ($H > 0$) unless the NEC is violated. 
There is a loophole, however, for the case of the closed Universe ($\kappa = +1$), where the bounce is possible provided $\rho$ and $p$ grow slower than $a^{-2}$ during the contraction phase (i.e. $\omega =p/\rho < - 1/3$)~\cite{Murphy:1973zz,Starobinsky,Durrer:1995mz,Falciano:2008gt}.

As for the Genesis scenario it is also ruled out in GR by the Penrose theorem and the NEC. Indeed, the expansion starting from the asymptotically Minkowski space implies the evolution from $H=0$ to positive values, hence, 
$\dot{H} > 0$, which according to eq.~\eqref{eq:dotH} requires NEC violation.
Alternatively, the same conclusion follows
from the covariant conservation of the energy-momentum tensor $\nabla_{\mu} T^{\mu\nu} = 0$ written for a homogeneous, isotropic Universe:
\[
\label{eq:conservation}
\frac{\mbox{d}\rho}{\mbox{d}t} = - 3 H (\rho + p).
\]
While the NEC is satisfied eq.~\eqref{eq:conservation} implies that the phase with the increasing energy density ($\dot{\rho} > 0$), which is crucial for the Genesis stage, is impossible in the expanding Universe. Hence, the Penrose theorem states that the cosmological expansion started from a singularity with an infinite energy density and infinite expansion rate. 

Therefore, violating the NEC~\eqref{eq:NEC} in a healthy way is a primary task for any non-singular cosmological scenario, which involves either a contraction phase before the Big Bang or an expansion phase starting from the asymptotically static empty space. 
While there exist different approaches to invoke the NEC violation, 
in this review we focus on the case when the latter is achieved by
modifying GR, which is 
realised within the scalar-tensor theories with higher-derivative Lagrangians, namely, Horndeski theories~\cite{Horndeski:1974wa} and their extensions.

%%%%%%%%%%%%%%%%%%%%%%%%%%%%%%%%%%
%%%%%%%%%%%%%%%%%%%%%%%%%%%%%%%%%%
\subsection{Horndeski theories as a suitable NEC-violating framework}
\label{sec:STT}

Scalar-tensor theories embody probably the simplest way to extend GR, which amounts to introducing an additional scalar degree of freedom on top of the two tensor modes already present within the original GR.
Even though the UV behaviour of the scalar-tensor theories is not much better than that of GR, the scalar-tensor family is of particular significance since most IR modifications of gravity 
can be described by scalar-tensor theories. 

The most general scalar-tensor theory with the second order equations of motion was formulated by Horndeski~\cite{Horndeski:1974wa}. The Horndeski theories are parametrized by four arbitrary functions $F$, $K$, $G_4$ and $G_5$
\footnote{There is no complete agreement on notations in literature and, hence, the functions $F$ and $K$ can be alternatively denoted as $G_2$ and $G_3$, respectively.}
of the scalar field $\pi$ and its kinetic term 
$X=g^{\mu\nu}\partial_{\mu}\pi \partial_{\nu}\pi $, and have the following form of the Lagrangian (metric signature $(+,-,-,-)$):
\begin{eqnarray}
  \label{eq:lagrangianH}
   &&
    \phantom{\mathcal{L}_2=F(\pi,X)\quad}
  % \mathcal{L}
  S_H = \int d^4x \sqrt{-g}\Big( \mathcal{L}_2 + \mathcal{L}_3 + \mathcal{L}_4 + \mathcal{L}_5
   % + \mathcal{L_{BH}}
   \Big),\\
  \label{L2H}
  &&\mathcal{L}_2=F(\pi,X),\\
  \label{L3H}
  &&\mathcal{L}_3=K(\pi,X)\Box\pi,\\
  \label{L4H}
  &&\mathcal{L}_4=-G_4(\pi,X)R+2G_{4X}(\pi,X)\left[\left(\Box\pi\right)^2-\pi_{;\mu\nu}\pi^{;\mu\nu}\right],\\
  \label{L5H}
  &&\mathcal{L}_5=G_5(\pi,X)G^{\mu\nu}\pi_{;\mu\nu}+\frac{1}{3}G_{5X}\left[\left(\Box\pi\right)^3-3\Box\pi\pi_{;\mu\nu}\pi^{;\mu\nu}+2\pi_{;\mu\nu}\pi^{;\mu\rho}\pi_{;\rho}^{\;\;\nu}\right],
\end{eqnarray}
where 
$R$ in eq.~\eqref{L4H} and $G^{\mu\nu}$ in eq.~\eqref{L5H} stand for the
Ricci scalar and Einstein tensor, respectively, and $\pi_{;\mu\nu}=\triangledown_\nu\triangledown_\mu\pi$,
$\Box\pi = g^{\mu\nu}\triangledown_\nu\triangledown_\mu\pi$,
$G_{iX}=\partial G_i/\partial X$. In fact, the form of the action~\eqref{eq:lagrangianH} is different from the original one formulated by Horndeski in~\cite{Horndeski:1974wa}
and was independly derived within a context of the Galileon theory~\cite{Nicolis:2008in} studies and its generalizations~\cite{Deffayet:2009wt,Deffayet:2009mn,Deffayet:2011gz}. Both forms are equivalent though and describe the same theory~\cite{Kobayashi:2011nu}. 
One of the main advantages of Horndeski theories is that by definition every scalar-tensor theory, whose equations of motion are second order, belong to the Horndeski group and can be cast into the form~\eqref{eq:lagrangianH}. In particular, upon a specific choice of
scalar potentials $F$, $K$, $G_4$ and $G_5$ one can reproduce, for example,
Einstein-Hilbert action, Brans-Dicke theory~\cite{Brans:1961sx}, $f(R)$-gravity theories~\cite{Sotiriou:2008rp,DeFelice:2010aj}, k-essence~\cite{Chiba:1999ka,Armendariz-Picon:2000nqq}, kinetic gravity braiding~\cite{Deffayet:2010qz,Pujolas:2011he}, Fab Four~\cite{Charmousis:2011ea,Charmousis:2011bf} (which involve couplings to the Gauss-Bonnet term), 
Dirac-Born-Infeld (DBI) galileons~\cite{deRham:2010eu,Goon:2011qf},
any inflationary theory based on a scalar field, e.g. k-inflation~\cite{Armendariz-Picon:1999hyi} and G-inflation~\cite{Kobayashi:2010cm}, 
and many more
(see e.g.~\cite{Heisenberg:2018vsk,KobayashiReview} for a thorough review).

 Higher-derivative terms in the Lagrangian~\eqref{eq:lagrangianH} are the hallmark of Horndeski theories. Since the corresponding equations of motion are still of the second order in derivatives, Horndeski theories are manifestly free of Ostrogradsky ghosts. However, the second order field equations provide a sufficient, but not necessary conditions for avoiding the Ostrogradsky ghost. Indeed, if the kinetic matrix of the system is degenerate, then, despite having equations of motions 
 with time-derivatives of order higher than two, 
 the extra degree of freedom (DOF) associated with the additional initial conditions
 gets eliminated. This is the observation which underlies the formulation of Degenerate Higher-Order Scalar-Tensor (DHOST) theories~\cite{Langlois:2015cwa,Langlois:2015skt,Crisostomi:2016czh,BenAchour:2016fzp}, which generalize Horndeski theories. The general form of the corresponding Lagrangian for both the quadratic and cubic subclasses
 \footnote{Here we adopt the standard naming of the subclasses based on the highest power of $\partial^2\pi$ involved.}
 of DHOST theories
can be found e.g. in a review~\cite{Langlois:2018dxi}.

The important difference between the DHOST theories and the Horndeski subclass is that the former are described by the third order equations of motion while still propagating $2+1$ DOFs. Another peculiar and quite useful feature is that different healthy subclasses of DHOST theories are related by disformal transformations~\cite{Crisostomi:2016czh,BenAchour:2016cay, Langlois:2018dxi}.
A conformal-disformal transformation~\cite{Bekenstein:1992pj} of a metric 
generally depends on two arbitrary functions $\Omega(\pi, X)$ and $\Gamma(\pi,X)$:
\[
\label{eq:disformal}
g_{\mu\nu} \to \Omega^2(\pi, X) \; g_{\mu\nu} + \Gamma(\pi,X) \partial_{\mu}\pi\partial_{\nu}\pi.
\]
Upon applying the transformation to the metric in Horndeski theory~\eqref{eq:lagrangianH}, one can obtain different resulting theories based on the specific choice of $\Omega(\pi, X)$ and $\Gamma(\pi,X)$. In particular, if both functions depend only on $\pi$, i.e. $\Omega=\Omega(\pi)$ and $\Gamma=\Gamma(\pi)$, the resulting theory is still a Horndeski theory~\cite{Bettoni:2013diz,Crisostomi:2016tcp}. 
The next option is to allow $\Gamma=\Gamma(\pi,X)$ with still $X$-independent conformal factor $\Omega=\Omega(\pi)$. Then the resulting theory is the so-called ”beyond Horndeski theory” or GLPV theory -- the first example of DHOST theories discovered by applying the disformal transformation~\eqref{eq:disformal} to the Einstein-Hilbert action~\cite{Zumalacarregui:2013pma}. Beyond Horndeski theories were also formulated independently in~\cite{
Gleyzes:2014dya,Gao:2014soa,Gleyzes:2014qga}. The general form of beyond Horndeski Lagrangian involves additional contributions with two new functions $F_4(\pi,X)$ and $F_5(\pi,X)$ along with those in eq.~\eqref{eq:lagrangianH}:  
\[
\label{eq:lagrangianBH}
% \mathcal{L}_{BH}=
S_{BH} = \int d^4x \sqrt{-g}
\left(
F_4(\pi,X)\epsilon^{\mu\nu\rho}_{\quad\;\sigma}\epsilon^{\mu'\nu'\rho'\sigma}\pi_{,\mu}\pi_{,\mu'}\pi_{;\nu\nu'}\pi_{;\rho\rho'}+
   % \\\nonumber&&\qquad+
   F_5(\pi{},X)\epsilon^{\mu\nu\rho\sigma}\epsilon^{\mu'\nu'\rho'\sigma'}\pi_{,\mu}\pi_{,\mu'}\pi_{;\nu\nu'}\pi_{;\rho\rho'}\pi_{;\sigma\sigma'}\right),
\]
where $\epsilon_{\mu\nu\rho\sigma}$ is the totally antisymmetric Levi-Civita tensor. It should be noted that not all combinations of terms in eqs.~\eqref{eq:lagrangianH} and~\eqref{eq:lagrangianBH} are allowed, since in some cases the additional propagating mode do arise in fact, which signals that the corresponding kinetic matrix is non-degenerate.  
The degeneracy requirement for the resulting theory imposes the following constraint on the Lagrangian functions $F_4$ and $F_5$:
\begin{equation}
\label{eq:F4F5relation}
{F}_4 \; G_{5X} X = -3 F_5 \;\left[{G}_4 - 2 X {G}_{4X} + \frac12 G_{5\pi} X\right],
\end{equation}
which was discussed in~\cite{Langlois:2015cwa,Crisostomi:2016tcp,BenAchour:2016fzp}, given without derivation in e.g.~\cite{Creminelli:2017sry}
and explicitly derived in e.g.~\cite{Mironov:2022ffa}. 
This relation for $F_4$ and $F_5$ can be viewed equivalently as follows: 
upon the disformal transformation of Horndeski theory with $\Omega(\pi)$ and $\Gamma(\pi,X)$, the resulting beyond Horndeski theory features the contributions of $F_4$ and $F_5$ which are compliant with the constraint~\eqref{eq:F4F5relation}.

Finally, upon applying the disformal transformation with both conformal and disformal factors depending on $X$, i.e. $\Omega(\pi, X)$ and $\Gamma(\pi,X)$, the Horndeski theory gets converted into the so-called ''Ia'' subclass of DHOST theory, which we specify below. The general from of the action for the quadratic DHOST theory reads
\footnote{We refrain from writing down the cubic DHOST action for brevity. As mentioned before, the complete action can be read off in e.g.~\cite{Langlois:2018dxi}.}:
\[
\label{eq:DHOSTquadr}
S_{DHOST} = \int d^4x \sqrt{-g}
\left( F(\pi,X) + K(\pi,X)\Box\pi + F_2(\pi,X) R + 
\sum_{i=1}^5 A_i(\pi,X) L_i \right),
\]
with
\[
L_1 = \pi_{;\mu\nu} \pi^{;\mu\nu}, \quad
L_2= \left(\Box{\pi}\right)^2, \quad
L_3= \pi^{,\mu} \pi_{;\mu\nu} \pi^{,\nu} \Box{\pi}, \quad
L_4= \pi^{,\mu} \pi_{;\mu\nu} \pi^{;\nu\rho} \pi_{,\rho}, \quad
L_5= \left( \pi^{,\mu} \pi_{;\mu\nu} \pi^{,\nu} \right)^2,
\]
where $F$, $K$, $F_2(\pi,X)$ and $A_i(\pi,X)$ are the functions analogous to $G_i$ in eq.~\eqref{eq:lagrangianH}, although interrelated by three equations, which implement the degeneracy constraints to preserve $2+1$ DOFs in the system (see e.g.~\cite{Langlois:2015cwa,Crisostomi:2016czh}).
DHOST theories are stable w.r.t. general conformal-disformal transformations~\eqref{eq:disformal}, meaning that the degeneracy of the kinetic matrix is preserved in result of such metric transformations~\cite{BenAchour:2016cay}. 
In what follows, we will concentrate on the
DHOST Ia subclass, which
is characterized by the following relations between the Lagrangian functions $F_2(\pi,X)$ and $A_i(\pi,X)$ (see e.g.~\cite{Langlois:2018dxi}):
\begin{subequations}
  \label{eq:DHOST_Ia}
  \begin{align}
A_2 &= - A_1\,,\\
\label{a4_A}
A_4&= \frac{1}{8(F_2-X A_1)^2}\left[-16 X A_1^3+4 (3F_2+16 XF_{2X})A_1^2
-(16X^2 F_{2X}-12XF_2) A_3 A_1
\right.
\cr
&\qquad 
\left.
-X^2 F_2 A_3^2-16 F_{2X}(3F_2+4XF_{2X})A_1+8F_2 (XF_{2X}-F_2)A_3+48F_2 F_{2X}^2\right],
\\
\label{a5_A}
A_5&=\frac{\left(4F_{2X}-2A_1+XA_3\right)\left(-2A_1^2-3XA_1A_3+4F_{2X} A_1+4F_2 A_3\right)}{8(F_2-XA_1)^2}\,.
  \end{align}
  \end{subequations}
DHOST Ia subclass involves both Horndeski and beyond Horndeski theories as special cases. The latter is supported by the fact, that
the result of the disformal transformation of Horndeski theory with $\Omega(\pi, X)$ and $\Gamma(\pi,X)$ corresponds to DHOST Ia subclass. This can be viewed alternatively as a characteristic feature of DHOST Ia theories: the Lagrangian~\eqref{eq:DHOST_Ia} can be mapped into the Horndeski form via the conformal-disformal transformation~\cite{BenAchour:2016cay}.

DHOST Ia subclass is of particular significance for cosmology, since the rest of subclasses are phenomenologically disfavoured (they either suffer from instabilities or feature the non-dynamical tensor modes~\cite{deRham:2016wji,Langlois:2017mxy}). Therefore, only the DHOST theories which are disformally related to Horndeski are viable.
Let us note that the two theories related by conformal-disformal transformation~\eqref{eq:disformal} are manifestly equivalent as long as the transformation is invertible and there is no additional matter components present in the system. We address the subtleties related to preserving the degeneracy upon coupling of additional matter to DHOST theories and other related issues in Sec.~\ref{sec:matter_coupling}.

As it is discussed and explicitly shown in e.g.~\cite{RubakovNEC, KobayashiReview} Horndeski theories and their extensions are capable of violating the NEC
\footnote{In the general case, when the gravity gets modified (this is the case for the Horndeski theory and its generalisations) the NEC~\eqref{eq:NEC} is replaced with the Null Convergence Condition (NCC)~\cite{Tipler:1978zz}: $R_{\mu\nu} n^{\mu}n^{\nu} > 0$, where $R_{\mu\nu}$ is Ricci tensor.}
and preserving the stability at the level of perturbations at the same time, provided one considers $\mathcal{L}_3$ subclass~\eqref{L3H} and higher. However, as we will see in Sec.~\ref{sec:nogo} not all subclasses of Horndeski theories~\eqref{eq:lagrangianH} are suitable for constructing non-singular cosmological models that are free from pathological degrees of freedom throughout the entire evolution of the Universe.

%%%%%%%%%%%%%%%%%%%%%%%%%%%%%%%%%%%%%%%%%%%%%%%%%%%%%%%%%%%%%%%%%%%%%
%%%%%%%%%%%%%%%%%%%%%%%%%%%%%%%%%%%%%%%%%%%%%%%%%%%%%%%%%%%%%%%%%%%%%
\section{Stability of a cosmological background in Horndeski theories and beyond}
\label{sec:stability}

Let us start with a brief revision of the linearized beyond Horndeski theory in a cosmological setting, which provides the necessary tools for the analysis of perturbations' behaviour. Our primary objective is to derive the stability criteria for a homogeneous and isotropic cosmological model.
In what follows we adhere to the covariant formalism, while the results of this section were also formulated within the effective field theory (EFT) approach~\cite{Creminelli:2016zwa,Cai:2016thi}.

Linear perturbations over the cosmological FLRW background can be classified
in a standard way according to their behaviour w.r.t. rotations into scalar, vector and tensor modes (see e.g.~\cite{Gorbunov:2011zzc}). In what follows we focus on the scalar and tensor perturbations, since the vector modes are non-dynamical in scalar-tensor theories.

We adopt the following parametrization for the metric perturbations:
\begin{equation}
\label{eq:FLRW_perturbed}
\mathrm{d}s^2 = N^2 \mathrm{d}t^2 - \gamma_{ij}(\mathrm{d}x^i+ N^i \mathrm{d}t)(\mathrm{d}x^j+N^j \mathrm{d}t),
\end{equation}
where 
\[
\label{eq:ADM}
N = 1+\alpha, \qquad N_i = \partial_i\beta, \qquad
\gamma_{ij}= a^2(t) e^{2\zeta} \left(\delta_{ij} + h_{ij}^T +
\dfrac12 h_{ik}^T {h^{k\:T}_j}\right).
\]
In eq.~\eqref{eq:ADM} $\alpha$, $\beta$ and $\zeta$ belong to the 
scalar sector of perturbations, while $h_{ik}^T$ denote transverse
traceless tensor modes ($h_{ii}^T = 0, \partial_i h_{ij}^T = 0$). 
Another contribution to the scalar sector comes from
the perturbations of the homogeneous scalar field $\pi(t)$, which are denoted as $\delta\pi$.
Note that in eq.~\eqref{eq:ADM} we have already removed the longitudinal component $\partial_i\partial_j E$
in $\gamma_{ij}$ by partially fixing the gauge. The residual gauge freedom is given by the following transformations:
\[
\label{eq:gauge}
\alpha \to \alpha + \dot{\xi}_0,\quad \beta \to \beta - \xi_0, \quad \delta\pi \to
\delta\pi + \xi_0\dot{\pi},\quad \zeta \to \zeta + \xi_0 H,
\]
where $H$ is the Hubble parameter, $\xi_0$ is the gauge function and as before an overdot stands for the time-derivative.
%%%

Substituting the metric~\eqref{eq:FLRW_perturbed} into the action 
$S_H + S_{BH}$,
see eqs.~\eqref{eq:lagrangianH} and~\eqref{eq:lagrangianBH},
and expanding it to the second order in perturbations, we obtain
$S^{(2)} = S^{(2)}_{tensor} + S^{(2)}_{scalar}$, where
\[
\label{eq:quadratic_action_tensor}
S^{(2)}_{tensor}=\frac18\int\mathrm{d}t\mathrm{d}^3x \;a^3
\left[{\mathcal{{G}_T}}\left(\dot{h}^T_{ij}\right)^2
-\dfrac{\mathcal{F_T}}{a^2}\left(\triangledown h_{ij}^T\right)^2
\right],
\]
with 
\begin{eqnarray}
\label{eq:GT_coeff_setup}
\mathcal{G_T}&=&2G_4-4G_{4X}X+G_{5\pi}X-2HG_{5X}X\dot{\pi} + 2F_4X^2+6HF_5X^2\dot{\pi},
\\
\label{eq:FT_coeff_setup}
\mathcal{F_T}&=&2G_4-2G_{5X}X\ddot{\pi}-G_{5\pi}X.
\end{eqnarray}

As for the scalar sector and the corresponding quadratic action $S^{(2)}_{scalar}$, its explicit form depends on the residual gauge fixing choice in eq.~\eqref{eq:gauge}. 
One of the standard approaches is to remove the perturbations of the scalar field $\delta\pi = 0$ (the unitary gauge). 
Then the quadratic action for the scalar sector reads
\footnote{We note that here and in what follows we make use of the background equations of motion, whose explicit form may be found in e.g.~\cite{Kolevatov:2017voe}.}:
\begin{equation}
\label{eq:quadratic_action_scalar}
\begin{aligned}
S^{(2)}_{scalar}=\int\mathrm{d}t\mathrm{d}^3x a^3\Bigg[
-3\mathcal{{G}_T}\dot{\zeta}^2+\mathcal{F_T}\dfrac{(\triangledown\zeta)^2}{a^2}
-2(\mathcal{G_T}+\cD \dot{\pi})\alpha\dfrac{\triangle\zeta}{a^2}
+2\mathcal{{G}_T}\dot{\zeta}\dfrac{\triangle\beta}{a^2}+6\Theta\alpha\dot{\zeta}-2\Theta\alpha\dfrac{\triangle\beta}{a^2}+\Sigma\alpha^2\Bigg],
\end{aligned}
\end{equation}
where $(\triangledown\zeta)^2 = \delta^{ij} \partial_i \zeta \partial_j \zeta$,
$\triangle = \delta^{ij} \partial_i \partial_j$, while $\cG_{\cT}$, $\cF_{\cT}$ are given in~\eqref{eq:GT_coeff_setup}-\eqref{eq:FT_coeff_setup}
and
\begin{eqnarray}
\label{eq:D_BH}
\mathcal{D}&=&-2F_4X\dot{\pi}-6HF_5X^2,
\\
\label{eq:Theta_coeff_setup}
\Theta&=&
-K_XX\dot{\pi}+2G_4H-8HG_{4X}X-8HG_{4XX}X^2+G_{4\pi}\dot{\pi}+2G_{4\pi X}X\dot{\pi}-5H^2G_{5X}X\dot{\pi}\nonumber\\
&&-2H^2G_{5XX}X^2\dot{\pi}+3HG_{5\pi}X+2HG_{5\pi X}X^2
% \\\nonumber&
+10HF_4X^2+4HF_{4X}X^3+21H^2F_5X^2\dot{\pi}
,\qquad\\
\label{eq:Sigma_coeff_setup}
\Sigma&=&F_XX+2F_{XX}X^2+12HK_XX\dot{\pi}+6HK_{XX}X^2\dot{\pi}-K_{\pi}X-K_{\pi X}X^2-6H^2G_4
\nonumber\\
% \nonumber
&&+42H^2G_{4X}X+96H^2G_{4XX}X^2+24H^2G_{4XXX}X^3-6HG_{4\pi}\dot{\pi}-30HG_{4\pi X}X\dot{\pi}
\nonumber\\
 \nonumber
&&-12HG_{4\pi XX}X^2\dot{\pi}+30H^3G_{5X}X\dot{\pi}+26H^3G_{5XX}X^2\dot{\pi}+4H^3G_{5XXX}X^3\dot{\pi}-18H^2G_{5\pi}X\\
 \nonumber
&&-27H^2G_{5\pi X}X^2-6H^2G_{5\pi XX}X^3-90H^2F_4X^2-78H^2F_{4X}X^3-12H^2F_{4XX}X^4\\
% \nonumber
&&-168H^3F_5X^2\dot{\pi}-102H^3F_{5X}X^3\dot{\pi}-12H^3F_{5XX}X^4\dot{\pi}.
\end{eqnarray}
The actions~\eqref{eq:quadratic_action_tensor} and~\eqref{eq:quadratic_action_scalar} were derived for the general Horndeski theory in~\cite{Kobayashi:2011nu} and later generalized to beyond Horndeski class e.g. in~\cite{Kolevatov:2017voe}.

The action~\eqref{eq:quadratic_action_scalar} does not involve time-derivatives of either $\alpha$ or $\beta$ and, hence, the variation of the action w.r.t. $\alpha$ and $\beta$ provides two constraint equations:
\begin{subequations}
\label{eq:constr_setup}
\begin{align}
\dfrac{\triangle\beta}{a^2}:&\qquad \alpha=\dfrac{\mathcal{{G}_T}\dot{\zeta}}{\Theta},\\
\alpha:&\qquad\dfrac{\triangle\beta}{a^2}=\dfrac{1}{\Theta}\left(3\Theta\dot{\zeta}-(\mathcal{G_T}+\cD \dot{\pi})\dfrac{\triangle\zeta}{a^2}+\Sigma\alpha\right).
\end{align}
\end{subequations}
Making use of these constraints to express $\alpha$ and $\triangle\beta$, we rewrite the action~\eqref{eq:quadratic_action_scalar} in terms of one dynamical DOF, which is the curvature perturbation $\zeta$:
\begin{equation}
\label{eq:unconstrained_action_scalar}
S^{(2)}_{\zeta}=\int\mathrm{d}t\mathrm{d}^3x a^3\left[\mathcal{G_S}\dot{\zeta}^2-\mathcal{F_S}\dfrac{(\triangledown\zeta)^2}{a^2}\right],
\end{equation}
where the following notations are introduced:
\begin{eqnarray}
\label{eq:GS_setup}
&&\mathcal{G_S}=\dfrac{\Sigma\mathcal{{G}_T}^2}{\Theta^2}+3\mathcal{{G}_T},\\
\label{eq:FS_setup}
&&\mathcal{F_S}=\dfrac{1}{a}\dfrac{\mathrm{d}\xi}{\mathrm{d}t}-\mathcal{F_T},\\
\label{eq:xi_func_setup}
&&\xi
=\dfrac{a\left(\mathcal{{G}_T}+\mathcal{D}\dot{\pi}\right)\mathcal{{G}_T}}{\Theta}.
\end{eqnarray}
The quadratic actions~\eqref{eq:quadratic_action_tensor} and~\eqref{eq:unconstrained_action_scalar}
describe one scalar ($\zeta$) and two tensor ($h^T_{ij}$) DOFs.
The propagation speed squared of
the scalar and tensor perturbations reads,
respectively:
\[
\label{eq:speeds}
c_\mathcal{T}^2=\dfrac{\mathcal{F_T}}{\mathcal{{G}_T}},\qquad c_\mathcal{S}^2=\dfrac{\mathcal{F_S}}{\mathcal{G_S}}.
\]

Let us briefly comment on the main
types of instabilities, 
which can possibly
arise at the linearized level of perturbations (see~\cite{RubakovNEC} for details). 
In the case of a homogeneous and isotropic
background, the coefficients $\mathcal{G_T}$, $\mathcal{F_T}$, 
$\mathcal{G_S}$ and $\mathcal{F_S}$ are the functions of time.
The most dangerous instabilities are those arising
in the high energy regime, i.e. when the characteristic scales of temporal
and spatial variations of $\zeta$ and $h^T_{ij}$ are considerably
smaller than that of the homogeneous background.
In the high energy
approximation, the coefficients  
$\mathcal{G_{S,T}}$ and $\mathcal{F_{S,T}}$ can be treated as
time-independent at relevant time intervals.
Then the following situations are possible 
(the notations $\mathcal{G_{S,T}}$, $\mathcal{F_{S,T}}$ refer
to pairs of coefficients $\mathcal{G_{S}}$, $\mathcal{F_{S}}$ or 
$\mathcal{G_{T}}$, $\mathcal{F_{T}}$):
\begin{itemize}
\item[(1)] Gradient instabilities (exponential growth 
of perturbations):
\[
\mathcal{G_{S,T}} >0, \quad \mathcal{F_{S,T}} < 0, \quad \mbox{or}  \quad \mathcal{G_{S,T}} < 0, \quad \mathcal{F_{S,T}} > 0.
\]
\item[(2)] Ghosts (catastrophic instability of the
  vacuum state, see Ref.~\cite{RubakovNEC} for discussion):
\[
 \mathcal{{G_{S,T}}} < 0, \quad \mathcal{{F_{S,T}}} < 0.
\]
\item[(3)] Stable solution:
\[
\label{eq:healthy}
\mathcal{{G_{S,T}}} >0, \quad \mathcal{{F_{S,T}}} >0.
\]
\end{itemize}
Let us note that we do not analyze the tachyonic instabilities,
{i.e. the modes with the negative mass or imaginary frequencies at low momenta,} since this case is likely to require performing the full-fledged stability analysis.

Another potential issue is related to superluminal propagation. 
Here and in what follows we choose the tactics of constraining superluminal modes from appearance.
It should be noted, however, that the superluminality and stability problems are not directly related to each other: superluminality may occur in a vicinity of a stable part of solution, see e.g.~\cite{Creminelli:2010ba,Dobre:2017pnt}.
So, the superluminality issue requires separate analysis and
in order to ensure that the speeds of both tensor and scalar modes~\eqref{eq:speeds} are luminal at most, the stability conditions in eq.~\eqref{eq:healthy} have to be modified as follows:
\begin{equation}
\label{eq:stability_cond_sublum}
\mathcal{{G}_T} \geq \mathcal{F_T} >0,\quad \mathcal{G_S}\geq\mathcal{F_S} > 0 \; .
\end{equation}
Hence, satisfying the set of constraints~\eqref{eq:stability_cond_sublum} allows one to guarantee that the cosmological solution in (beyond) Horndeski theories is stable and does not admit the propagation of superluminal modes.
We address the issue of superluminalities in further details in Sec.~\ref{sec:matter_coupling_superlum}.

In result, within different Horndeski subclasses there was suggested a great number of non-singular cosmological settings with either a bouncing dynamics in e.g.~\cite{Qiu:2011cy,Easson:2011zy,Cai:2012va,Osipov:2013ssa,Koehn:2013upa,Battarra:2014tga,Qiu:2015nha,Ijjas:2016tpn,Ijjas:2016vtq,Sberna:2017xqv} or Genesis epoch~\cite{Creminelli:2006xe,Creminelli:2010ba,Liu:2011ns,Hinterbichler:2012mv,Liu:2012ww,Hinterbichler:2012fr,Creminelli:2012my,Hinterbichler:2012yn,Liu:2013xt,Rubakov:2013kaa,Elder:2013gya,Nishi:2014bsa,Pirtskhalava:2014esa,Nishi:2015pta,Kobayashi:2015gga,Nishi:2016wty,Nishi:2016ljg}, which comply with the above stability conditions at least throughout and around the NEC-violating phase.

Although (beyond) Horndeski theories are suitable for non-singular cosmologies 
with no obvious pathologies 
% around and 
during the NEC-violating stage, ensuring that the solution is free from instabilities during the entire evolution turned out to be a non-trivial task. 
The problem with global or ''complete'' stability of non-singular cosmological scenarios was first addressed in $\cL_3$ subclass of Horndeski theory~\eqref{L3H} and formulated in the form of the no-go theorem in~\cite{LMR}, and later generalized for the case of general Horndeski theory~\eqref{eq:lagrangianH} in~\cite{KobayashiNogo} (see also~\cite{Mironov:2019fop}).
In the following section we revisit the no-go theorem 
for the completely stable non-singular solutions of FLRW type in Horndeski theory and then discuss the existing ways to circumvent it.

%%%%%%%%%%%%%%%%%%%%%%%%%%%%%%%%%%%%%%%%%%%%%%%%%%%%%%%%%%%%%%%%%%%%%
\section{Stability of non-singular cosmologies: the no-go theorem and ways to circumvent it}
\label{sec:nogo}

\subsection{The no-go theorem in Horndeski theory}
\label{sec:theorem}

Let us start with setting the terms and specifying the notion of a completely stable solution in FLRW background. Namely, in what follows ''complete stability'' implies that the pathological perturbations do not develop around the background solution during the entire period of evolution in time. 
According to eq.~\eqref{eq:healthy}, the absence of ghosts and
gradient instabilities as well as superluminal propagation over a homogeneous background solution
suggests the following restrictions on the coefficients 
in the quadratic actions~\eqref{eq:quadratic_action_tensor} and~\eqref{eq:unconstrained_action_scalar}: 
\begin{equation}
\label{eq:stability_cond_nogo}
\mathcal{{G}_T} \geq \mathcal{F_T}> \epsilon >0,\quad \mathcal{G_S}\geq\mathcal{F_S} >\epsilon> 0 \; .
\end{equation}
Hereafter $\epsilon$ denotes a positive constant,
whose actual value is irrelevant
for our reasoning, so it may vary in different formulas below.
The reason behind introducing
this constant in eq.~\eqref{eq:stability_cond_nogo} is
to avoid the situations when $\mathcal{G_{S,T}},\mathcal{F_{S,T}} \to 0$
in the asymptotic past and/or future,
which we address separately in Sec.~\ref{sec:strong_coupling}. In what follows we focus on non-singular cosmological solutions only with 
$a > \epsilon > 0$, where $\epsilon$ ensures that the scale factor is bounded from below and, hence, guarantees the geodesic completeness
(we comment on the case of a quickly vanishing scale factor in the asymptotic past or future in Sec.~\ref{sec:strong_coupling}).

The no-go theorem states that there are no completely stable non-singular cosmological solutions with $a > \epsilon > 0$ in Horndeski theory, since the instability inevitably shows up in the scalar sector of perturbations provided one considers the entire evolution of the solution. 

We prove the theorem by contradiction and set off with the assumption that there exists a non-singular cosmology in Horndeski theory which complies with the stability constraints~\eqref{eq:stability_cond_nogo} at all times. 
Let us now analyse
the constraint for the gradient coefficient $\cF_{\cS}$ in the quadratic action for scalar modes, see eqs.~\eqref{eq:unconstrained_action_scalar} and~\eqref{eq:FS_setup}-\eqref{eq:xi_func_setup}:
\[
\label{eq:nogoFS}
\dfrac{\mathrm{d}\xi}{\mathrm{d}t} =
a \left(\mathcal{F_S} +\mathcal{F_T}\right) > \epsilon > 0,
\]
with
\[
\label{eq:xi_H}
\xi=\dfrac{a\mathcal{G_T}^2}{\Theta}.
\]
Recall that $\cD = 0$ for Horndeski theory (see eq.~\eqref{eq:D_BH}).
According to the stability criteria~\eqref{eq:stability_cond_nogo} and the definition~\eqref{eq:nogoFS} ${\mathrm{d}\xi}/{\mathrm{d}t}$ is strictly positive at all times, hence, $\xi$ is a monotonous function of time. 
Let us now show that this monotonous function $\xi$ inevitably crosses zero at some moment of time, which is in direct contradiction with the definition of $\xi$ in eq.~\eqref{eq:xi_H}, since $\cG_\cT > \epsilon > 0$, $a > \epsilon > 0$
and $\Theta$ is a smooth function of time, finite at all points (see eq.~\eqref{eq:Theta_coeff_setup} for definition).

We start with integrating~\eqref{eq:nogoFS} over time from $t_i$ to some $t_f > t_i$ and get:
\begin{equation}
\label{eq:no-go_integral}
\int^{t_f}_{t_i}{a\left(\cF_\cS+\cF_\cT\right)\mathrm{d}t}  =\xi(t_f)-\xi(t_i).
\end{equation}
It is natural to assume that the integral on the left-hand side is divergent at time infinities,
i.e. as $t_i \to -\infty$ and $t_f \to +\infty$ (the contrary would require both $\cF_\cT$ and $\cF_\cS$ to approach zero sufficiently fast in the asymptotic past and/or future, which is in contradiction with~\eqref{eq:stability_cond_nogo}). 
Now, suppose that $\xi(t_i) < 0$, then 
\[
\int^{t_f}_{t_i}{a\left(\cF_\cS+\cF_\cT\right)\mathrm{d}t} - |\xi(t_i)|  = \xi(t_f),
\]
and since the integral is divergent in the asymptotic future, there exists the sufficiently large $t_f$ so that $\xi(t_f) > 0$. Hence, based on the continuity of the monotonous function, $\xi(t)$ crosses zero at some moment between $t_i$ and $t_f$. 
However, as we pointed out above, according to the definition of $\xi$~\eqref{eq:xi_H} and constraints~\eqref{eq:stability_cond_nogo} the latter is impossible for non-singular cosmologies with $a > \epsilon > 0$. The only option for $\xi$ to cross zero is when $\Theta \to \infty$, which corresponds to a singularity in the classical solution, since $\Theta$ is a combination of Lagrangian functions and background values, see eq.~\eqref{eq:Theta_coeff_setup}. Therefore, we deduce that $\xi$ has to be positive at all times. But upon writing
\[
\xi(t_i) = \xi(t_f) - \int^{t_f}_{t_i}{a\left(\cF_\cS+\cF_\cT\right)\mathrm{d}t}  ,
\]
we find that the right-hand side becomes negative for $t_i \to -\infty$, hence,
$\xi(t_i) < 0$. This contradiction proves the theorem. 

The no-go theorem was proved to hold even with the additional matter present along with the Horndeski scalar~\cite{KobayashiNogo,Kolevatov:2016ppi,Akama:2017jsa} as well as in the case of the open Universe~\cite{Akama:2018cqv}.

Let us now discuss the existing loopholes and possible ways to evade the no-go theorem. In practice, 
all strategies of getting around the no-go theorem suggest the ways of resolving the contradiction between the fact that by definition $\xi$~\eqref{eq:xi_H} cannot cross zero in a healthy way 
but at the same time this zero-crossing is necessary to avoid gradient instabilities~\eqref{eq:nogoFS}.

%%%%%%%%%%%%%%%%%%%%%%%%%%%%%%%%%%%%%%%%%%%%%%%%%%%%%%%%%%%%%%%%%%%%%
% \subsection{Vanishing coefficients and potential strong coupling}
\subsection{Evading the no-go theorem: non-zero $\xi$ and potential strong coupling}
\label{sec:strong_coupling}

One of the ways to resolve the contradiction is to protect $\xi$ from crossing zero.
This loophole is related to relaxing the assumption that the central integral in~\eqref{eq:no-go_integral} is divergent as $t_i \to -\infty$ and/or $t_f \to +\infty$, which naturally followed from the requirement 
$\cF_\cT, \cF_\cS > \epsilon > 0$ in eq.~\eqref{eq:stability_cond_nogo}. 
As it was 
was already noted in~\cite{KobayashiNogo}, one may allow sufficiently rapid decay
$\mathcal{G_{S,T}},\mathcal{F_{S,T}} \to 0$ in asymptotic past and/or future, so that
the integral in~\eqref{eq:no-go_integral} is convergent as $t_i \to -\infty$ and/or $t_f \to +\infty$ and, hence the function $\xi$ does not necessarily cross zero. This loophole was utilized in~\cite{KobayashiNogo,Ijjas:2016vtq,Nishi:2016ljg} in construction of the specific Genesis scenarios and the bouncing Universe in Horndeski theory without the apparent gradient instability. 

It should be noted that this approach is tricky since at least naively $\mathcal{G_{S,T}},\mathcal{F_{S,T}} \to 0$ signals that a strong coupling regime takes place in the asymptotic past and/or future. 
However, it was explicitly shown 
in~\cite{Ageeva:2018lko,Ageeva:2020gti,Ageeva:2021yik,Akama:2022usl} 
that there exists a region in parameter space of Horndeski Lagrangian where the problem of strong coupling does not occur and the completely stable non-singular solution can be constructed in Horndeski theory.

Another way to make the integral in~\eqref{eq:no-go_integral} convergent is to abandon the requirement for the scale factor to be strictly non-vanishing and allow asymptotically $a \to 0$. This is exactly the strategy that was suggested in~\cite{LMR} within the modified Genesis scenario. It should be noted that the case of the convergent integral suffer from geodesic incompleteness~\cite{Creminelli:2016zwa}, which is not necessarily pathological but requires an additional study~\cite{Rubakov:2022fqk}.

%%%%%%%%%%%%%%%%%%%%%%%%%%%%%%%%%%%%%%%%%%%%%%%%%%%%%%%%%%%%%%%%%%%%%
\subsection{Evading the no-go theorem: divergent $\xi$ and $\gamma$-crossing}
\label{sec:gamma_crossing}

An alternative approach to evading the no-go theorem is to make $\xi$ in~\eqref{eq:xi_H} cross zero by adjusting the behaviour of the coefficient $\Theta$.
There is an option of allowing $\Theta$ and $\cG_\cT$ to vanish simultaneously, i.e. $\Theta(t_*)=0$ and $\cG_\cT(t_*)=0$, so that $\xi$ could cross zero at the moment $t_*$. However, this option implies fine-tuning and 
faces the problem of strong coupling, as $\cG_{\cT}$ vanishes in a finite moment of time $t_*$.

Let us note that the option of $\Theta=0$ at any moment of time might seem generally unacceptable without fine-tuning due to the fact, that $\Theta$ is involved in denominators of both $\cG_\cS$ and $\cF_\cS$~\eqref{eq:GS_setup}-\eqref{eq:FS_setup}. What makes things look even worse is that $\Theta$ is present in denominators of both constraints~\eqref{eq:constr_setup}, which questions the validity of the derived quadratic action for the scalar mode~\eqref{eq:unconstrained_action_scalar}. 
These problems related to $\Theta$ crossing zero were addressed in~\cite{Mironov:2018oec} where the moment of $\Theta=0$ was referred to as $\gamma$-crossing (the naming is motivated by~\cite{Ijjas:2016vtq,Ijjas:2017pei,Dobre:2017pnt}, where
the issue was initially addressed in different notations, i.e. $\Theta \equiv \gamma$).
It was explicitly shown that all perturbation variables including the non-dynamical $\alpha$ and $\beta$ in eqs.~\eqref{eq:constr_setup} are in fact regular even at $\gamma$-crossing and no other divergencies occur, hence, $\Theta=0$ is totally acceptable. 
Let us stress that $\gamma$-crossing itself does not help to evade the no-go theorem, but this phenomenon turns out to be
crucial for constructing cosmological solutions with bounce and Genesis, whose asymptotics are simple as $t \to \pm \infty$, i.e. where the theory reduces to GR.

A peculiar approach for evading the no-go theorem inspired by $\gamma$-crossing was suggested in~\cite{Mironov:2022quk}, where $\Theta$ was identically vanishing, i.e. $\Theta \equiv 0$ at all times. This case implies imposing a specific constraint on the Lagrangian functions and a
different strategy of removing non-dynamical variables $\alpha$ and $\beta$ in the quadratic action~\eqref{eq:quadratic_action_scalar}. It was shown that in the resulting Horndeski theory there are no dynamical scalar DOF, hence, the theory propagates only 2 DOFs like the GR. {However, allowing anisotropy in the cosmological setting
revives
the scalar DOF in this type of constrained Horndeski subclass, and immediately invokes the instability~\cite{Mironov:2023nzz}.}

\subsection{Evading the no-go theorem: the cuscuton}
\label{sec:cuscuton}

Yet another special subclasses of Horndeski theory that allow one to evade the no-go theorem are given by the
cuscuton~\cite{Afshordi:2006ad,Afshordi:2007yx} and 
the extended cuscuton~\cite{Afshordi:2009tt,Iyonaga:2018vnu} theories. The underlying idea is
quite similar to that with $\Theta\equiv 0$ and amounts to constraining the Lagrangian functions in a special manner so
that the scalar DOF becomes non-dynamical. 
Originally this option was realized within a subclass of k-essence,
where an additional relation is imposed on the Lagrangian functions, see eqs.~\eqref{eq:lagrangianH}: $F_X X + 2 F_{XX} X^2 =0$. 
Indeed, this constraint follows from the requirement 
\[
\label{eq:sigma_cuscuton}
\Sigma = - \frac{3 \Theta^2}{\cG_\mathcal{T}},
\] 
which gives $\Sigma = -6H^2 G_4$ based on the
definitions of $\Theta$ and $\cG_{\cT}$
in eqs.~\eqref{eq:GT_coeff_setup} and~\eqref{eq:Theta_coeff_setup} for the k-essence case. 
Upon the comparison of this result with the definition of $\Sigma$~\eqref{eq:Sigma_coeff_setup} one comes to the constraint for the Lagrangian function in question.
Most importantly, this specific choice in eq.~\eqref{eq:sigma_cuscuton} results in $\cG_\mathcal{S}=0$ (see eq.~\eqref{eq:GS_setup}) and, hence, the scalar mode
is indeed non-dynamical~\cite{Afshordi:2007yx}.
This type of constraint was further generalized to the case of the general Horndeski and even beyond Horndeski theories~\cite{Iyonaga:2018vnu}. 
Being a singular case of Horndeski theory, just like the option with $\Theta\equiv 0$ above, the cuscuton theory and its {additional degeneracy condition}~\eqref{eq:sigma_cuscuton} are more robust since, for example, 
% it is easy to see, that 
the anisotropy in a cosmological background does not affect the number of dynamical DOFs.
Moreover, the Hamiltonian analysis for this type of theories shows that above any cosmological background there is no propagating scalar mode~\cite{Gomes:2017tzd,Iyonaga:2018vnu}. This result relies on the unitary {slicing} (i.e. the scalar field $\pi(t)$ is time-dependent only) which can always be chosen for the everywhere time-like gradient of a scalar field. {However, in a general configuration of the scalar field adopting the unitary slicing is not possible and the propagating mode reappears and might be pathological.} {This situation resembles the defining feature of the U-DHOST theories~\cite{DeFelice:2018ewo}: these theories seem to be degenerate when written in the unitary slicing, but 
there is an apparent extra DOF
as soon as one considers their fully covariant version with no specific slicing or gauge choice.}

In result, there were suggested several specific examples of the bouncing solutions within the cuscuton theory, see e.g.~\cite{Boruah:2018pvq,Kim:2020iwq,HosseiniMansoori:2022xnq,Ganz:2022zgs}.

\subsection{Evading the no-go theorem: going beyond Horndeski theory}
\label{sec:BH}

Finally, to circumvent the no-go theorem it is possible to extend the Horndeski theory and consider more general scalar-tensor theories. 
In particular, one may consider the more general geometry with torsion in Horndeski-Cartan theory~\cite{Mironov:2023kzt,Mironov:2023wxn,Mironov:2024ffx} or teleparallel Horndeski gravity~\cite{Ahmedov:2023num,Ahmedov:2023lot}.
The conventional option, however, is to consider beyond Horndeski theory~\eqref{eq:lagrangianBH} or DHOST theories~\eqref{eq:DHOSTquadr}. Indeed, already in beyond Horndeski theory the definition of $\xi$ is modified as compared to the general Horndeski case, since the coefficient $\cD \neq 0$ in~\eqref{eq:xi_func_setup} due to the non-trivial functions $F_4(\pi,X)$ 
and $F_5(\pi,X)$. While the coefficient $\cG_\cT$ is still responsible for stability in the tensor sector and has to be always positive, the combination $(\cG +\cD \dot{\pi})$ can take any values including zero and negative ones. Hence, thanks to the unconstrained $\cD$ the variable $\xi$ can monotonously grow and cross zero at some moment of time.
In this case the no-go theorem no longer holds.

This opportunity to go beyond Horndeski was explored both within the 
EFT framework~\cite{Creminelli:2016zwa,Cai:2016thi,Cai:2017tku,Cai:2017dxl,Ye:2019frg,Ye:2019sth} 
and in the covariant 
language~\cite{Cai:2017dyi,Kolevatov:2017voe,Mironov:2018oec,Mironov:2019qjt,Mironov:2019haz,Ilyas:2020qja,Ilyas:2020zcb,Zhu:2021whu,Zhu:2021ggm}, 
where explicit examples of stable non-singular cosmologies with the bouncing or Genesis stages were put forward.

%%%%%%%%%%%%%%%%%%%%%%%%%%%%%%%%%%%%%%%%%%%%%%%%%%%%%%%%%%%%%%%%%%%%%
\subsection{The no-go theorem and disformal transformations}
\label{sec:no-go_disformal}

Evading the no-go theorem by extending the theory to beyond Horndeski class~\eqref{eq:lagrangianBH} might seem somewhat contradictory in view of the disformal relation between Horndeski and beyond Horndeski theories mentioned above and discussed in detail in e.g.~\cite{Gleyzes:2014qga,Crisostomi:2016czh,Crisostomi:2016tcp}. Indeed, the disformal transformation~\eqref{eq:disformal} is a field redefinition, and while it is invertible it cannot change the number of DOFs~\cite{Arroja:2015wpa,Domenech:2015tca,Takahashi:2017zgr} and cannot affect stability of the solution.

Therefore, the natural question is how is it possible that beyond Horndeski theory admits completely stable non-singular solutions while general Horndeski theory does not? The answer is
that the disformal transformation, which turns beyond Horndeski Lagrangian admitting stable solution into general Horndeski form, becomes singular at some moment of time. This result was first obtained within the EFT approach in~\cite{Creminelli:2016zwa} and later confirmed in the covariant formalism~\cite{Mironov:2019haz,Mironov:2022ffa}. 
In a nutshell,
the finding is based on the form of transformation rules for $X$-derivatives of the Lagrangian functions 
$G_4$ and $G_5$
under
disformal transformation~\eqref{eq:disformal} with $\Omega(\pi,X) = 1$:
\begin{eqnarray}
\label{eq:G4XG5X}
&\bar{G}_{4\bar{X}} \equiv \dfrac{\partial \bar{G}_4}{\partial \bar{X}}  = \left( \hat{G}_4 (1+\Gamma X)- \dfrac{1}{2} \hat{G}_{4}(\Gamma + X \Gamma_X) \right)\dfrac{\sqrt{1+\Gamma X}}{1-\Gamma_X X^2}, \\
\nonumber
&\bar{G}_{5\bar{X}} \equiv \dfrac{\partial \bar{G}_5}{\partial \bar{X}} =
G_{5X} \dfrac{(1+\Gamma X)^{5/2}}{1 - \Gamma_X X^2},
\end{eqnarray}
where the functions $\bar{G}_{4\bar{X}}$
and $\bar{G}_{5\bar{X}}$ belong to Horndeski theory, while $\hat{G}_{4}$
\footnote{Here we preserve the notations of~\cite{Mironov:2022ffa} where the hat over $\hat{G}_4$ emphasises that this contribution originates from disformal transformation of $\bar{G}_4$, but not $\bar{G}_5$.}
and $G_{5X}$ are their counterparts from the beyond Horndeski Lagrangian. 
Both transformation rules involve the same denominator, which can be rewritten as follows (see~\cite{Mironov:2022ffa} for details):
\begin{equation}
\label{eq:gamma_x_frac}
\dfrac{1}{1- \Gamma_X X^2}= \dfrac{\mathcal{G_T}}{\mathcal{G_T} + \cal{D}\dot{\pi}}\;,
\end{equation}
where $\cG_{\cT}$ and $\cD$ are similar to those involved in the definition of $\xi$ in eq.~\eqref{eq:xi_func_setup}. 
As discussed in Sec.~\ref{sec:BH} the combination $(\mathcal{G_T} + \cal{D}\dot{\pi})$ is supposed to cross zero at some moment, so that $\xi$ could cross zero in a healthy way  
and satisfy the stability constraint~\eqref{eq:nogoFS} at any time.
We see that
that the denominator of the transformation rules~\eqref{eq:G4XG5X} goes through zero at the same moment as $\xi$ does. This means that both Lagrangian functions $\bar{G}_{4\bar{X}}$ and $\bar{G}_{5\bar{X}}$ diverge at that moment of time. This proves that once there is a completely stable non-singular cosmological model in beyond Horndeski theory, the latter theory cannot be disformally transformed into the Horndeski class due to singularity in the corresponding transformation rules. So in fact there is no contradiction between the no-go theorem and the existence of completely stable cosmologies in seemingly disformally related theories.

As mentioned in Sec.~\ref{sec:STT} the {invertible} conformal-disformal transformation of the general form~\eqref{eq:disformal} connects the Horndeski, beyond Horndeski and DHOST Ia theories. 
Lately there were certain developments in this direction and the new classes of scalar-tensor theories were discovered with the help of even more general transformations of a metric. 
In particular, there were put forward
the generalized disformal Horndeski (GDH) theories that are connected to Horndeski via an invertible generalized disformal transformation~\cite{Babichev:2019twf,Babichev:2021bim,Takahashi:2021ttd,Takahashi:2022mew} of the following form:
\[
\label{eq:double_disformal}
{g}_{\mu\nu} \to \bar{F}_0 g_{\mu\nu}+\bar{F}_1\pi_\mu\pi_\nu+\bar{F}_2(\pi_\mu X_\nu+\pi_\nu X_\mu)+\bar{F}_3X_\mu X_\nu\, ,
\]
with functions $\bar{F}_i = \bar{F}_i (\pi,X,Y,Z)$ ($i=0,1,2,3$), where $Y=\pi_\mu X_\mu$ and $Z=X_\mu X_\mu$, and which are constrained by the invertibility condition, imposed on the transformation~\eqref{eq:double_disformal} (see e.g.~\cite{Takahashi:2021ttd}). The Lagrangian of the GDH theories explicitly contains third derivatives of the scalar field, but the theory still propagates $2+1$ DOFs since it can be converted into Horndeski via an invertible field redefinition. The GDH theory naturally contains DHOST Ia subclass as a special case, but the theory is more general and has non-DHOST subclasses as well. Notably, the GDH theory does not involve other subclasses of DHOST theory, which are not disformally related to Horndeski theory, however, those extra subclasses do not have viable cosmological application, so we do not discuss them in this review.

It turns out that the disformal transformations can be generalized even further, so that still more general scalar-tensor theories can be constructed. There is a possibility to include arbitrarily high derivatives of the scalar field into the disformal transformation~\cite{Takahashi:2023vva}
or even metric derivatives~\cite{Babichev:2024eoh}.

%%%%%%%%%%%%%%%%%%%%%%%%%%%%%%%%%%%%%%%%%%%%%%%%%%%%%%%%%%%%%%%%%%%%%
\section{Stability of non-singular cosmological models and matter coupling}
\label{sec:matter_coupling}

This whole machinery related to the disformal transformations and the degenerate theories perfectly works for the pure scalar-tensor theories.
However, different problems often arise in theories beyond Horndeski subclass when the coupling to other matter fields is introduced, while the Horndeski theory itself is safe in this regard.

In this section we discuss two types of problems related to matter coupling. The first one has to do with the fact, that the introduction of an additional set of fields can spoil the degeneracy conditions for the scalar sector, which protects the system form the Ostrogradsky ghost. 
The second problem is the appearance of a superluminally propagating mode in the scalar sector, which is the case when the matter with a sound speed close to the speed of light interacts even minimally with the gravitational sector within the scalar-tensor theory.

\subsection{Matter coupling and potentially ruined degeneracy}
\label{sec:matter_coupling_degeneracy}

The issue of the spoiled degeneracy upon matter coupling can be investigated in two different ways. First, one may study the spectrum of the joint theory, which consists of the general scalar-tensor part and the additional matter, and find out if the Ostrogradsky mode is present. 
The alternative way, which underlies the more common approach, is to analyse the theory, which is disformally related to the original one. In particular, since all the phenomenologically promising scalar-tensor theories are related to Horndeski via the invertible (general) disformal transformations, 
one can make use of this feature and disformally reduce any generalization of Horndeski theory with an additional matter into the Horndeski subclass, where 
the additional matter is coupled to the disformally related metric, {i.e. the coupling is no longer minimal in terms of the original metric.}

A detailed study showed that the matter which is coupled to the standard disformal metric~\eqref{eq:disformal} does not spoil the degeneracy of the Horndeski theory~\cite{Deffayet:2020ypa,Garcia-Saenz:2021pxb}. In other words, both beyond Horndeski and DHOST Ia theories allow healthy matter coupling. This result holds if the matter does not feature higher derivatives itself (i.e. does not belong to Horndeski type) and if the {coupling terms} do not involve derivatives. So, for example, couplings to a conventional scalar field and to the classical fermions are allowed. The {counterexamples} are given by the 
lower subclasses of Horndeski theory~\eqref{L3H} (and higher), 
which exemplify the matter that involves higher derivatives, and the non-gauge invariant vector field, which is coupled through the derivatives -- in both cases the resulting kinetic matrix is not degenerate anymore.
In the
more general theories like GDH the situation gets more complicated: unless one adopts the unitary {slicing} the only subclass that preserves its degeneracy upon the coupling to scalars and fermions is the DHOST Ia theory
itself~\cite{Takahashi:2022ctx,Ikeda:2023ntu}.

{Let us note that the reasoning above is formulated with no specific choice of a background and implies that the invertible disformal transformations do exist at all moments. The latter is not the case for a completely stable non-singular cosmological models.
Indeed, in Sec.~\ref{sec:no-go_disformal} it was shown that the beyond Horndeski theory, which admits a completely stable non-singular solution, cannot be disformally reduced into Horndeski subclass due to the existence of singular point(s) in the
corresponding
disformal transformation rules. Therefore, in this case the 
analysis based on disformal transformations cannot be used  
to check the degeneracy of beyond Horndeski theory with additional matter
in a finite number of points. This indicates that the theory's degeneracy in these singular points has to be studied separately (e.g. by analysing the spectrum of the linearized theory). 
}

{Moreover,}
the discussion above is relevant for the scalar-tensor theories with $2+1$ DOFs coupled to the additional matter. However, the situation is somewhat more subtle
within 
{"double degenerate"}
cases of such scalar-tensor theories, where the "first" degeneracy in the kinetic matrix removes the would be Ostrogradsky ghost, while the "second" one makes the scalar mode non-dynamical 
, like in e.g. the cuscuton theory (see Sec.~\ref{sec:cuscuton}).
This kind of special subclasses of scalar-tensor theories are more susceptible to the appearance of a new propagating mode after matter coupling, since the additional degeneracy condition (like~\eqref{eq:sigma_cuscuton} in the case of the cuscuton) is ruined even by the minimal coupling.
The latter was explicitly shown for the case of 'veiled' gravity theory~\cite{Deffayet:2020ypa}, but other cases like the cuscuton may also be problematic in this sense.

%%%%%%%%%%%%%%%%%%%%%%%%%%%%%%%%%%%%%%%%%%%%%%%%%%%%%%%%%%%%%%%%%%%%%
\subsection{Matter coupling and potentially induced superluminality}
\label{sec:matter_coupling_superlum}

Let us now revisit the issue of superluminal propagation which was briefly mentioned while formulating the set of stability constraints in Sec.~\ref{sec:stability}, see eq.~\eqref{eq:stability_cond_sublum}. Even though, as discussed above, the beyond Horndeski and DHOST theories remain degenerate when coupled to an additional matter, hence, protected from the Ostrogradsky ghost, the properties of the coupled matter itself could be altered: in particular, the corresponding sound speed may exceed that of light. 

The potential
appearance of superluminalities was shown to be a characteristic feature of the scalar-tensor theories like Horndeski family, see e.g.~\cite{Adams:2006sv,Bruneton:2006gf,Kang:2007vs,Babichev:2007dw,Nicolis:2008in,Burrage:2011cr,Evslin:2011rj,Hinterbichler:2009kq,deFromont:2013iwa,Kolevatov:2015iqa}.
Generally, the  modes propagating at superluminal speed over an arbitrary background solution are considered undesirable, since their existence indicates that the theory cannot descend, as a low energy effective field theory, from any UV-complete, Lorentz-covariant theory~\cite{Adams:2006sv}. However, one often does not claim to define the Lagrangian in the entire phase space: 
it is usually adequate to keep only those terms in the Lagrangian, which are sufficient for developing the solution and analyzing its stability (i.e., the terms that do not vanish on the given solution and its close neighborhood). In that case, the minimal requirement is the absence of superluminality for perturbations about the cosmological solution of interest and in its vicinity. In particular, this way of treating the superluminalities was discussed in both Genesis and bouncing scenarios, see e.g.~\cite{Creminelli:2010ba,Dobre:2017pnt,Creminelli:2012my,Mironov:2019mye,Easson:2013bda}.

Another tricky point is related to the way how coupling of the additional matter to Horndeski theories and their generalizations affects the propagations speed of perturbations over a cosmological background. For example, it was explicitly shown for the Genesis model in Horndeski theory~\cite{Creminelli:2012my}
that upon adding of a whatever tiny amount of an external matter (ideal fluid) with a normal equation of state ($0 < w < 1$)
the superluminal mode appears in some otherwise healthy region of a phase space~\cite{Easson:2013bda}. 
The relation between the emergent superluminality and the coupling of an additional matter is based on the kinetic mixing or braiding of the Horndeski scalar with metric~\cite{Pujolas:2011he}, which implies that the sound speed of the Horndeski scalar changes in the presence of other matter components. 

Things were shown to be different once one goes beyond Horndeski: e.g. in
~\cite{Mironov:2019mye} it was proved that a specifically designed beyond Horndeski Lagrangian, which on its own admits a stable and subluminal bouncing solution, remains free of superluminalities upon adding extra matter in the form of perfect fluid with the equation of state parameter $w \leq 1/3$ (or even somewhat larger). 
However, upon analysing the sound speeds of scalar modes over a cosmological background in the general setting ''beyond Horndeski + perfect fluid'', it was found that for $w$ equal or close to 1, one of the scalar propagation speeds inevitably becomes superluminal (this does not necessarily happen for $w$ substantially smaller than 1)~\cite{Mironov:2019mye}. This effect has to do with the fact, already noticed in~\cite{Gleyzes:2014dya,Gleyzes:2014qga,Langlois:2017mxy,Crisostomi:2018bsp}, that due to the specific structure of beyond Horndeski Lagrangian, there is kinetic mixing between matter and scalar field perturbations, and hence the sound speeds of both scalar modes get modified (the superluminal one is predominantly the sound wave in matter). This finding of the emergent superluminality has been supported by similar result in the case where instead of a perfect fluid, a conventional, minimally coupled scalar field (whose propagation is luminal, $c_m = 1$) is added to a cosmological setup in beyond Horndeski theory~\cite{Mironov:2020mfo} and also generalized for the DHOST Ia case~\cite{Mironov:2020pqh}.

In the following sections we explicitly demonstrate how the propagation speeds of the scalar perturbations are interrelated in the joint system of DHOST Ia theory and a minimally coupled additional
scalar field of the most general type.

%%%%%%%%%%%%%%%%%%%%%%%%%%%%%%%%%%%%%%%%%%%%%%%%%%%%%%%%%%%%%%%%%%%%%
\subsection{DHOST Ia with the extra scalar: the emergent superluminality}

First, we briefly discuss the quadratic action for perturbations within pure DHOST Ia subclass~\eqref{eq:DHOSTquadr}-\eqref{eq:DHOST_Ia} and, hence, generalize the results for the linearized beyond Horndeski case given in Sec.~\ref{sec:stability}.
Then we introduce coupling of an additional matter to the DHOST Ia setup and discuss its impact on stability. In this section we closely follow the reasoning of~\cite{Mironov:2020pqh}.

Our starting point is the (perturbed)
spatially flat FLRW background~\eqref{eq:FLRW_perturbed} 
with {a} rolling DHOST field $\pi=\pi (t)$.
In full analogy to Sec.~\ref{sec:stability} we consider  
the unitary gauge
$\delta \pi =0$, so that the
{dynamical}
perturbations 
in the DHOST sector are still given by the tensor modes
$h_{ij}^T$ and the scalar mode 
$\tilde{\zeta}$ (which differs from the original $\zeta$ in Sec.~\ref{sec:stability} and is explicitly defined below).
In pure DHOST~Ia theory, the unconstrained second
order action reads
\footnote{Here we adopt the notations for the coefficients similar to those in beyond Horndeski case, see eqs.~\eqref{eq:quadratic_action_tensor} and~\eqref{eq:unconstrained_action_scalar}. }
\begin{equation}
\label{eq:quadr_action_DHOSTIa-noscalar}
\begin{aligned}
  S^{(2)}_{\pi}=
  \int\mathrm{d}t\,\mathrm{d}^3x \,a^3
  \Bigg[\left(\dfrac{\mathcal{{G}_T}}{8}\left(\dot{h}^T_{ik}\right)^2-
    \dfrac{\mathcal{F_T}}{8a^2}\left(\partial_i h_{kl}^T\right)^2\right) +
 \left( \mathcal{{G}_S} \dot{\tilde{\zeta}}^2
- \dfrac{1}{a^2}  \mathcal{{F}_S} (\partial_i \tilde{\zeta})^2 \right)
\Bigg] \; ,
\end{aligned}
\end{equation}
where
\begin{subequations}
\label{eq:App_list_of_coefficients_1}
\begin{align}
 \mathcal{G_T} &= -2 F_2 + 2 A_1 X, \label{eq:App_Gt}\\ 
 \mathcal{F_T} &= -2 F_2,\label{eq:App_Ft}
  \\
  \mathcal{G_S} & = \dfrac{\tilde{\Sigma}\mathcal{{G}_T}^2}{\tilde{\Theta}^2}
    +3\mathcal{{G}_T}, \label{eq:App_Gs}
  \\
  \mathcal{F_S} &= \dfrac{1}{a}\dfrac{\mathrm{d}}{\mathrm{d}t}
  \left[ \dfrac{a \;\mathcal{{G}_T}\left(\mathcal{G_T} + \mathcal{D} \dot{\pi} + \mathcal{F_T}\Delta\right)}{\tilde{\Theta}}\right]
  -\mathcal{F_T},
   \label{eq:App_Fs}
\end{align}
\end{subequations}
with {
\begin{subequations}
  \begin{align}
    & \mathcal{D} = -2 A_1 \dot {\pi} + 4 F_ {2 X} \dot {\pi},
    \label{eq:calD_DHOST}\\
&\Delta =  \frac{X}{2 \mathcal{G_T}}\left(
    2 A_1- 4 F_{2X} - A_3 X \right) \; ,\label{eq:Delta}\\
    %\\
&\tilde{\Theta} = \Theta - \mathcal{G_T}\dot{\Delta},
\label{eq:theta_DHOST} \\
&\tilde{\Sigma} = \Sigma + 3\mathcal{G_T} \dot{\Delta}^2+ 6\tilde{\Theta} \dot{\Delta} 
-\dfrac{3}{a^3}\dfrac{\mathrm{d}}{\mathrm{d}t}\Big[a^3\left(\tilde{\Theta} +\mathcal{G_T} \dot{\Delta}\right)\Delta\Big],
\label{eq:sigma_DHOST}
  \end{align}
\end{subequations}
where the expressions for $\Sigma$ and $\Theta$ are quite lengthy and are
given in Appendix for completeness.
In these notations the scalar perturbation is
$\tilde{\zeta} = \zeta - \alpha \Delta$. The result for beyond Horndeski case can be restored upon the following choice of Lagrangian functions~\cite{Langlois:2018dxi}: 
\[
\label{eq:DHOSTIa_BH_correspondense}
F_2 = -G_4, \qquad A_1 = -A_2 = -(2 G_{4X} - X F_4), 
\qquad A_3 = - A_4 =- 2 F_4, \qquad A_5 = 0.
\]
Note that $\Delta$ identically vanishes in this case and we get back to the original curvature perturbation $\tilde{\zeta} \equiv \zeta $.
We note in passing that
hereafter
we do not use the background equations of motion when deriving
the action for perturbations.

As before the stability of the background
requires  
$\mathcal{G_T},\mathcal{F_T},\mathcal{G_S},\mathcal{F_S}
  > \epsilon > 0$ in~\eqref{eq:quadr_action_DHOSTIa-noscalar},
   while the sound speeds squared are restricted to be luminal at most
  \[
  \label{eq:speeds_DHOSTIa}
  c_{\mathcal{S} , 0}^2 =  \frac{\mathcal{F_S}}{\mathcal{{G_S}}} \leq 1 \; , \;\;\;\;
  c_{\mathcal{T}}^2 =  \frac{\mathcal{F_T}}{\mathcal{{G_T}}} \leq 1\;.
  \]

Now let us turn to the additional matter component
and consider
another scalar field $\chi$ in the form of k-essence~\cite{Armendariz-Picon:2000ulo}:
\[
\label{eq:action_setup_kess}
S_{\chi} = \int\mathrm{d}^4x\sqrt{-g} \,P(\chi,Y), \quad Y = g^{\mu\nu}\chi_{,\mu}\chi_{,\nu}\,.
\]
The Lagrangian in eq.~\eqref{eq:action_setup_kess} describes 
a minimally coupled
scalar field $\chi$ of the most general type
{(assuming the absence of second derivatives in the
  Lagrangian)}. 
In this context k-essence successfully mimics the irrotational, barotropic
fluid. 
The field equation for the homogeneous scalar $\chi=\chi(t)$ in the flat FLRW setting reads: 
\[
\label{eq:background_kess}
\ddot{\chi} + 3 \dfrac{P_Y}{Q} H \dot{\chi} - \dfrac{P_{\chi} 
- 2 Y P_{\chi Y}}{2\,Q} = 0 \; ,
\]
where $Q = P_Y +2 Y P_{YY}$.
{In a spatially homogeneous background
  (possibly rolling, $Y=\dot{\chi}^2 \neq 0$), the stability conditions
  for the scalar field $\chi$ have a standard form
  \[
  P_Y > 0 \; , \;\;\;\;\;  Q >0 \; ,
  \label{apr24-20-1}
  \]
  while the propagation speed of perturbations equals to
  \[
\label{eq:cm}
c_m^2 = \dfrac{P_Y}{Q} \; .
\]
{Our main result on the emergent superluminality} in
what follows applies most straightforwardly to the conventional
scalar field with
\[
P(\chi,Y)= \dfrac{1}{2} Y - V(\chi) \; ,
\label{eq:conventional_scalar}
\]
and $c_m^2 = 1$, but for now we retain full generality and do not make
any assumptions on the form of the function $P(\chi, Y)$.
}

We now combine the two theories and
consider 
DHOST~Ia along with the extra scalar theory~\eqref{eq:action_setup_kess}
over
the background where both $\dot{\pi}$ and  $\dot{\chi}$ do not vanish
(in particular, $Y= \dot{\chi}^2\neq 0$). The expressions for
$\mathcal{{G}_T}$ and $\mathcal{F_T}$ in eqs.~\eqref{eq:App_Gt}-\eqref{eq:App_Ft} do not get modified,
so the tensor
perturbations remain (sub)luminal according to our requirement in~\eqref{eq:speeds_DHOSTIa}. 
On the contrary,
the situation in the scalar sector changes dramatically.
The non-vanishing background $\dot{\chi}$ induces mixing between
the scalars $\tilde{\zeta}$ and $\delta \chi$~\cite{Langlois:2017mxy}, 
so the {unconstrained} quadratic
action {in the joint scalar sector}
reads 
\begin{equation}
\label{eq:quadr_action_DHOSTIa+k}
\begin{aligned}
  S^{(2) \, scalar}_{\pi+\chi}
  =\int\mathrm{d}t\,\mathrm{d}^3x \,a^3\Bigg[
    G_{AB} \, \dot{v}^A \dot{v}^B 
- \dfrac{1}{a^2} F_{AB} \, \partial_i\,{v^A} \partial_i\,{v^B} + \dots
\Bigg]
,
\end{aligned}
\end{equation}
where $A,B=1,2$, 
$v^1 = \tz$, $v^2 = \delta\chi$,
and the dots stand for the terms with less than two derivatives. %For small $Y$
The matrices $G_{AB}$ and $F_{AB}$
have the following forms:
\[
\label{eq:GAB_FAB}
G_{AB} = 
\begin{pmatrix}
\mathcal{G_S} + \dfrac{\mathcal{{G}_T}^2}{\tilde{\Theta}^2} YQ 
& \dot{\chi} Q g
 \\ \\
 \dot{\chi} Q g
 & Q
\end{pmatrix}, ~~~~~~~~~
%\ee
%\be
F_{AB} = 
\begin{pmatrix}
\mathcal{F_S} 
&  \dot{\chi} P_Y f
\\ \\
 \dot{\chi} P_Y f & 
 P_Y
\end{pmatrix}\; ,
\]
where
\begin{subequations}
\label{eq:gf}
  \begin{align}
    g &= -\dfrac{\mathcal{{G}_T}}{\tilde{\Theta}}  \left(1-3\dfrac{P_Y}{Q}\Delta
    \right),
    \\
    f &=  -\dfrac{\left(\mathcal{G_T} + \mathcal{D} \dot{\pi} + \mathcal{F_T}
      \Delta
      \right)}{\tilde{\Theta}} \; .
  \end{align}
  \end{subequations}
Let us note that these expressions are valid for any $Y$.
The two sound speeds squared $c_{\mathcal{S} , \pm}^2$ are given by eigenvalues of the
matrix $G^{-1} F$, i.e., they satisfy
\[
\label{eq:det}
\mbox{det}\left(F_{AB} - c_{\mathcal{S}}^2 G_{AB}\right) = 0 \; .
\]
Before proceeding with the DHOST Ia case let us address
 the results for the (beyond) Horndeski theories~\cite{Mironov:2020mfo}.

%%%%%%%%%%%%%%%%%%%%%%%%%%%%%%%%%%%%%%%%%%%%%%%%%%%%%%
\subsubsection{Beyond Horndeski case: superluminality due to a conventional scalar field}
\label{sec:superlum_BH}

For the beyond Horndeski case both matrices $G_{AB}$ and $F_{AB}$ simplify since $\Delta \equiv 0$, see eqs.~\eqref{eq:Delta},~\eqref{eq:DHOSTIa_BH_correspondense} and~\eqref{eq:gf}.
Hence, the eq.~\eqref{eq:det} gives the following propagation speeds of $\tz \equiv \zeta$ and $\delta\chi$ (recall that $c_m^2 = P_Y/Q$):
\[
c_{\mathcal{S}\, \pm}^2 = \dfrac12 (c_m^2 + \mathcal{A})
\pm  \dfrac12 \sqrt{(c_m^2 - \mathcal{A})^2 + \mathcal{B}},
\label{eq:speeds_BH}
\]
where
\[
\mathcal{A}= \frac{\mathcal{F_S}}{\mathcal{G_S}} -
\frac{YP_Y}{\mathcal{G_S}} \, \frac{\mathcal{G_T}(\mathcal{G_T} + 2 \mathcal{D}\dot{\pi})}{ \Theta^2} \; , \;\;\;\;\;\;
\mathcal{B}= 4c_m^2\frac{YP_Y}{\mathcal{G_S}}
\frac{(\mathcal{D} \dot{\pi})^2}{ \Theta^2} \; .
  \nonumber
\]
In a stable and rolling background both matrices $G_{AB}$ and $F_{AB}$ must be positive definite {($G_{11}, G_{22}>0$, $\det G > 0$
  and $F_{11}, F_{22}>0$, $\det F > 0 $):
\[
\label{eq:stability_conditions_scalar_BH}
\mathcal{{G}_S} > 0 \; , 
\quad \mathcal{{F}_S}>0,  
\quad  P_Y>0 \;, \quad Q > 0\; , \quad \mathcal{{F}_S} - Y P_Y\dfrac{\left(\mathcal{G_T} + \mathcal{D} \dot{\pi}\right)^2}{\Theta^2} > 0,
\]
and, hence, with $Y>0$
    the coefficient   $\mathcal{B}$ is positive. 
This gives immediately 
    \[
    \label{eq:superlumBH}
    c_{\mathcal{S}\, +}^2 > c_m^2 \;\;\;\;\mbox{for}\;\; Y\neq 0 \; .
    \]
    So, if the propagation of the scalar perturbation
    $\delta\chi$
is luminal, $c_m=1$, then according to~\eqref{eq:superlumBH}
it becomes superluminal in the ``beyond Horndeski + scalar field''
system. 
The interpretation of $c^2_{\mathcal{S}\, +}$ as the speed of perturbations in the additional matter can be supported by restoration of the Horndeski limit.

Indeed,
in the unextended Horndeski case~\eqref{eq:lagrangianH} the coefficient $\cD$ vanishes (see eq.~\eqref{eq:D_BH}), 
so that $g=f$ in matrices $G_{AB}$ and $F_{AB}$ in\eqref{eq:GAB_FAB} 
and the matrix $G^{-1}F$ becomes triangular.
In result, the speed of $\delta\chi$ recovers its standard value
$c_m^2$, while the propagations speed of the curvature perturbation $\zeta$
is modified. Indeed, with $\cD = 0$
the propagations speeds in~\eqref{eq:speeds_BH} get simplified as follows
\[
\label{eq:speed_Horndeski}
{c_{\mathcal{S} \, -}^2}|_{\mathcal{D}=0}
= \dfrac{\mathcal{F_S}}{\mathcal{{G}_S}} -
\dfrac{Y P_Y}{\mathcal{{G}_S}} \dfrac{\mathcal{{G}_T}^2}{\Theta^2}, 
\quad 
{c_{\mathcal{S}\, +}^2}|_{\mathcal{D}=0} = c_m^2,
\]
where we explicitly see that ${c_{\mathcal{S}\, +}^2}$ 
belongs to $\delta\chi$. 

In particular, the result~~\eqref{eq:superlumBH} indicates that in the case of a conventional scalar field~\eqref{eq:conventional_scalar}, where $c_m^2 = 1$ for any $Y$, the scalar sector of ''beyond Horndeski + $P(\chi,Y)$'' system
features a superluminal mode even with a tiny amount of kinetic energy associated with a rolling scalar $\chi(t)$. Therefore, we conclude that the multi-scalar system with the minimal coupling of a conventional scalar field to beyond Horndeski theory is exposed to the emergent superluminality.

%%%%%%%%%%%%%%%%%%%%%%%%%%%%%%%%%%%%%%%%%%%%%%%%%%%%%%%%%%%%%%%%%%
\subsubsection{DHOST Ia case: superluminality and an exceptional subclass }
\label{sec:superlum_DHOST}

Now we get back to full-fledged DHOST Ia case 
and assume that the background is stable, and DHOST perturbations about it are not superluminal. The stability conditions for the scalar sector in the combined system of DHOST Ia and k-essence 
generalize those for ''beyond Horndeski theory+k--essence'' in~\eqref{eq:stability_conditions_scalar_BH}
and read explicitly:
\[
\begin{aligned}
\label{eq:stability_scalar_DHOST}
 \mathcal{{G}_S}>0,  \quad 
 \mathcal{{F}_S} > 0, \quad
&P_Y > 0, \quad
Q > 0, \\
\mathcal{{F}_S} - Y P_Y\dfrac{\left(\mathcal{G_T} + \mathcal{D} \dot{\pi}+ \mathcal{F_T}\Delta\right)^2}{\tilde{\Theta}^2} > 0, \quad
&1+ \dfrac{6\mathcal{G_T}^2}{\tilde{\Theta}^2} \dfrac{Y P_Y}{\mathcal{{G}_S}} \left(1-\dfrac32\dfrac{P_Y}{Q}\Delta\right) \Delta > 0,
\end{aligned}
\]
where
the second line
is characteristic of the 
combined theory, while the first line coincides with the 
stability conditions
for separated DHOST Ia theory and k-essence. Note that the beyond Horndeski case~\eqref{eq:stability_conditions_scalar_BH}{}
is straightforwardly restored upon taking $\Delta=0$.
At this point we neither  impose any
further constraints on the background 
nor {assume any relations
  other than \eqref{eq:DHOST_Ia} between}
  the Lagrangian functions in~\eqref{eq:DHOSTquadr} yet.

Let us consider the general situation with $f\neq g$ and take
$Y$ to be small in eqs.~\eqref{eq:GAB_FAB}-\eqref{eq:gf}. Then we distinguish two situations:

(i) If $c_m^2 \neq c_{\mathcal{S} , 0}^2$
(i.e., $\ P_Y/Q
\neq  \mathcal{F_S}/\mathcal{G_S}$),
then one of the sound speeds, following from eq.~\eqref{eq:det}, is $ c_{\mathcal{S} , -} =
c_{\mathcal{S} , 0} + \mathcal{O}(Y)$, while the other is given by
  \[
  c_{\mathcal{S} , +}^2 = c_m^2
  \left(1 + \frac{Y(f-g)^2}{\mathcal{G_S}(c_m^2 - c_{\mathcal{S} , 0}^2)}
  \right)  + \mathcal{O}(Y^2)
  \; .
  \]
  This means, in particular, that in the theory 
  of the conventional scalar field~\eqref{eq:conventional_scalar} with $c_m=1$ and subluminal
  DHOST~{Ia}
  with  $c_{\mathcal{S} , 0} < 1$, the mode which is predominantly
  $\delta \chi$ becomes superluminal at small but non-zero
  background values of $Y$. 

  (ii) For
  $c_m^2 = c_{\mathcal{S} , 0}^2 $, the sound speeds are given by
  \[
  c_{\mathcal{S} , \pm}^2  = c_m^2 \left[1 \pm
  \left(\frac{Y P_Y(f-g)^2}{\mathcal{G_S}} \right)^{1/2}\right]
+ \mathcal{O}(Y)
\; ,
  \]
  which again shows that  in the case
  of the conventional scalar field with $c_m=1$ one 
  of the modes becomes superluminal at small $Y$.

Let us finally turn to the special case $f=g$. By making use of eq.~\eqref{eq:GAB_FAB} one finds that 
in this case
the matrix $G^{-1}F$ is triangular  {\it for any
    value of $Y$}, so that
  one of the sound speeds
  remains unmodified,
  $ c_{\mathcal{S} , +}^2=c_m^2=P_Y/Q$,
  while another is not necessarily
  superluminal  ($ c_{\mathcal{S} , -} = c_{\mathcal{S} , 0} + \mathcal{O}(Y)$
  for small $Y$).
  For the luminal extra scalar with $c_m=1$, the condition $f=g$ provides a constraint
  $\mathcal{D}\dot{\pi} = - (3\mathcal{G_T}+\mathcal{F_T})\Delta$
    on the DHOST Lagrangian, or,
  explicitly,
  \[
A_3 = \dfrac{2(A_1-2F_{2X})(A_1 X - 2 F_2)}{X(3 A_1 X - 4 F_2)} \; .
\label{eq:special_DHOST}
\]
This is an exceptional subclass of DHOST~Ia theories in which
adding extra luminal scalar field does not necessarily leads to
superluminality. Note that this subclass includes the theory
with $A_1 = 2F_{2 X}$ and $A_3=0$, which is Horndeski, and does {\it not}
include beyond Horndeski (GLPV) theories with
$XA_3 = 2A_1 - 4 F_{2 X}$ (the latter relation is inconsistent with
\eqref{eq:special_DHOST} for $\mathcal{G_T} \neq 0$, 
{see eq.~\eqref{eq:App_Gt}}).
This observation is in agreement with the result for beyond Horndeski in Sec.~\ref{sec:superlum_BH}.

Thus, the emergent superluminality over the cosmological background generally takes place in the DHOST Ia class upon adding to the system of even a tiny amount of kinetic energy $Y>0$ related to the additional scalar field $\chi$~\eqref{eq:conventional_scalar} with $c_m^2 =1$. There is, however, an exceptional subclass of DHOST Ia theories given by~\eqref{eq:special_DHOST}, where the superluminality of perturbations in the scalar sector is not induced.

%%%%%%%%%%%%%%%%%%%%%%%%%%%%%%%%%%%%%%%%%%%%%%%%%%%%%%%%%%%%%%%%%%%%%%%%%%%%%%
\section{Conclusion}
\label{sec:conclusion}

In this mini-review we have briefly discussed the updates on the cosmological scenarios without the initial singularity in Horndeski theories and their generalizations.
The review is focused on
ensuring linear stability at the perturbation level
{w.r.t. ghosts and gradient instabilities}, which has been shown to pose significant challenges to the construction of non-singular cosmological models that are stable throughout the entire period of evolution. 
We have revisited the no-go theorem, which excludes the existence of completely stable non-singular solutions within Horndeski theory, and discussed in details the existing approaches to evade it, highlighting the specific subclasses of non-extended Horndeski theory where the no-go can be circumvented.
Furthermore, we addressed
the role of disformal transformations, which interconnect different healthy subclasses of Horndeski and generalized theories like DHOST Ia, and discussed 
the effect of additional matter coupling on the degeneracy of the joint field theory. We have also paid special attention to the potential appearance of superluminal modes in the ''DHOST Ia + extra matter'' system: it was shown that a superluminal scalar mode generally arises over the cosmological background 
in the DHOST Ia class upon adding to the system even a tiny amount of kinetic energy sourced by the additional scalar field with the sound speed equal to that of light. This effect of the emergent superluminality, however, does not take place for a specific DHOST Ia subclass, which was identified explicitly and includes Horndeski theory as a special case but does not include beyond Horndeski (GLPV) subclass.

Scalar-tensor theories like Horndeski theories and their generalizations constitute a particularly rich framework of modified gravity theories, which are suitable for creating and analyzing various cosmological scenarios, describing both the early and late-time Universe. 
In result of studying specific cosmological models like those without the initial singularity one might be hopeful to not only discover fundamental aspects behind the theories they are build in,
but also to gain further insights into theoretical aspects of gravity and physics of the early Universe.

%%%%%%%%%%%%%%%%%%%%%%%%%%%%%%%%%%%%%%%%%%%%%%%%%%%%%%%%%%%%%%%%%%%%%%%%%%%%%%%%%%%%
\section*{Acknowledgements}
A considerable part of this essay reviews the results of studies of non-singular cosmologies carried out by Valery Rubakov, both in independent work and in collaboration with his students and colleagues. Likewise, more recent research works of S. Mironov, V. Volkova, Y. Ageeva, P. Petrov, A. Shtennikova and M. Valencia-Villegas are still inspired by Prof. Rubakov's ideas, which were discussed with him on various occasions in the past.

The work on Sec.~\ref{sec:stability} and~\ref{sec:matter_coupling}
has been supported by Russian Science Foundation grant № 24-72-10110,\\
\href{https://rscf.ru/project/24-72-10110/}{  https://rscf.ru/project/24-72-10110/}. The work of S.M. and V.V. on Sec.~\ref{sec:nogo} was partially supported by Theoretical Physics and Mathematics Advancement Foundation “BASIS”.

%%%%%%%%%%%%%%%%%%%%%%%%%%%%%%%%%%%%%%%%%%%%%%%%%%%%%%%%%%%%%%%%%%%%%
\section*{Appendix}

Here we provide the explicit expressions for $\Theta$ and $\Sigma$ involved in the quadratic action for DHOST Ia theories, see eqs.~\eqref{eq:theta_DHOST}-\eqref{eq:sigma_DHOST}:
% \begin{eqnarray}
% \label{eq:App_list_of_coefficients}
 \begin{eqnarray}
\Theta &=&  (-A_{3X}+A_5) \ddot{\pi} \dot{\pi} X^2-\dot{\pi} (F_{2\pi}+\ddot{\pi}
(-3A_1+6 F_{2X}))-2 F_2 H + X (3 A_1+2 F_{2X}) H 
\\\nonumber
&&+ X^2 \left(-2 A_{1X} H + \frac{3}{2}
 (4 A_{1X}+A_3) H \right)
% \nonumber\\
% &&
-\dot{\pi} X \left( \ddot{\pi}
(- 2 A_{1X}+ \frac{3}{2} A_3- A_4+4 F_{2XX})+2 F_{2\pi X}+K_X \right),  
% \label{eq:App_Theta_action}
% \\
\end{eqnarray}
\begin{eqnarray}
\Sigma &=& 6 F_2 H^2-2 \ddot{\pi} (A_{5\pi} X+A_{5X X} \ddot{\pi}) \dot{\pi}^8
-\dot{\pi}^7 (2 A_{5X} \dddot{\pi}+3 A_{3\pi X} H+6 A_{5X} \ddot{\pi} H)
\\
&&+6 \dot{\pi} [ F_{2\pi} H-2 \ddot{\pi} (A_1-2 F_{2X}) H]+\dot{\pi}^6 [-2 (A_{3\pi X}+A_{4\pi X}+4 A_{5\pi}) \ddot{\pi}
  \nonumber\\
    &&-(2 A_{3X X}+2 A_{4X X}+13 {A_5X}) \ddot{\pi}^2-3 A_{3X} \dot{H}
  -3 (4 A_{1X X} + 3 A_{3X}) H^2]
\nonumber\\
  &&+\dot{\pi}^2 [-3 (A_3+A_4)
    \ddot{\pi}^2-12 \dot{H} (-A_1+2 F_{2X})+ F_{X} %\\&
    -42 F_{2X} H^2-K_{\pi}]
\nonumber\\
&& +\dot{\pi}^4 [-6 (A_{3\pi}+A_{4\pi})
    \ddot{\pi} 
    -(9 A_{3X}+9 A_{4X}+12 A_5) \ddot{\pi}^2-3 \dot{H} (-2 A_{1X}+3 A_3+4 F_{2X X})\nonumber\\
&& +
    2  F_{X X} -36 A_{1X} H^2-27 A_3 H^2-24 F_{2X X} H^2-K_{\pi X}]
-\dot{\pi}^3 \;\dddot{\pi} [6 (A_3+A_4) \dddot{\pi}
  \nonumber\\
  &&+3 \ddot{\pi} (2 A_{1X}+3 A_3+6 A_4-4 F_{2X X}) H+6 H (-2 A_{1\pi}-F_{2\pi X}-2 K_X)]
\nonumber\\\nonumber
&&-\dot{\pi}^5 \left[2 (A_{3X}+A_{4X}+4 A_5)
+3 (A_{3X}+2 A_{4X}+8 A_5) \ddot{\pi} H+3 H (-2 A_{1\pi X}+3 A_{3\pi}-2 K_{X X})\right], \label{eq:App_Sigma_action}
\end{eqnarray}
where the functions $A_4$ and $A_5$ are given by~\eqref{a4_A} and~\eqref{a5_A}, respectively.
% We note that the
% coefficients $\tilde{\Theta}$ and $\tilde{\Sigma}$ in eqs.~\eqref{eq:theta_DHOST}-\eqref{eq:sigma_DHOST} play similar roles as the
% coefficients $\Theta$ and $\Sigma$, respectively,
% in (beyond) Horndeski case.

%%%%%%%%%%%%%%%%%%%%%%%%%%%%%%%%%%%%%%%%%%%%%%%%%%%%%%%%%%%%%%%%%%%%%%%%%%%%%%%%%%%%

\end{document}